\documentclass{cjaa}                    

\usepackage{graphicx}                   
\input{epsf.sty}                        
\input{psfig.sty}                       

\def\NrPulsars{187 } 
\def\NrNewDrifters{42}
\def\NrDrifters{68 }
\def\NrPulsarsSNR{106 }

\def\DriftPercentageSNR{54}

\begin{document}

   \title{Statistics of the drifting subpulse phenomenon
}

   \volnopage{Vol.0 (200x) No.0, 000--000}      
   \setcounter{page}{1}          

   \author{P. Weltevrede
      \inst{1}\mailto{}
   \and R.~T. Edwards
      \inst{1,3}
   \and B.~W. Stappers
      \inst{2,1}
      }
   \offprints{P. Weltevrede}                   

   \institute{Astronomical Institute ``Anton Pannekoek'', 
        University of Amsterdam,\hspace{5cm}
        Kruislaan 403, 1098 SJ Amsterdam, The Netherlands \\
             \email{wltvrede@science.uva.nl}
        \and
             Stichting ASTRON, Postbus 2, 7990 AA Dwingeloo, The Netherlands
        \and
             CSIRO Australia Telescope National Facility, PO Box 76, Epping NSW 1710,  Australia\\
          }

   \date{Received~~2001 month day; accepted~~2001~~month day}

   \abstract{ We present the statistical results of a systematic,
unbiased search for subpulse modulation of \NrPulsars pulsars
performed with the Westerbork Synthesis Radio Telescope (WSRT) in the
Netherlands at an observing wavelength of 21 cm (Weltevrede
et~al.~\cite{wes06}). We have increased the list of pulsars that show
the drifting subpulse phenomenon by \NrNewDrifters, indicating that
more than 55\% of the pulsars that show this phenomenon. The large
number of new drifters we have found allows us, for the first time, to
do meaningful statistics on the drifting phenomenon. We find that the
drifting phenomenon is correlated with the pulsar age such that
drifting is more likely to occur in older pulsars. Pulsars that drift
more coherently seem to be older and have a lower modulation
index. Contrary claims from older studies, both $P_3$ (the repetition
period of the drifting subpulse pattern) and the drift direction are
found to be uncorrelated with other pulsar parameters.
 }

   \authorrunning{}            
   \titlerunning{}  

   \maketitle

%
%
\section{Introduction}           
\label{sect:intro}
\begin{figure}[t]
\begin{center}
\begin{minipage}{.48\textwidth}
\rotatebox{270}{\resizebox{1.38\hsize}{!}{\includegraphics[angle=0]{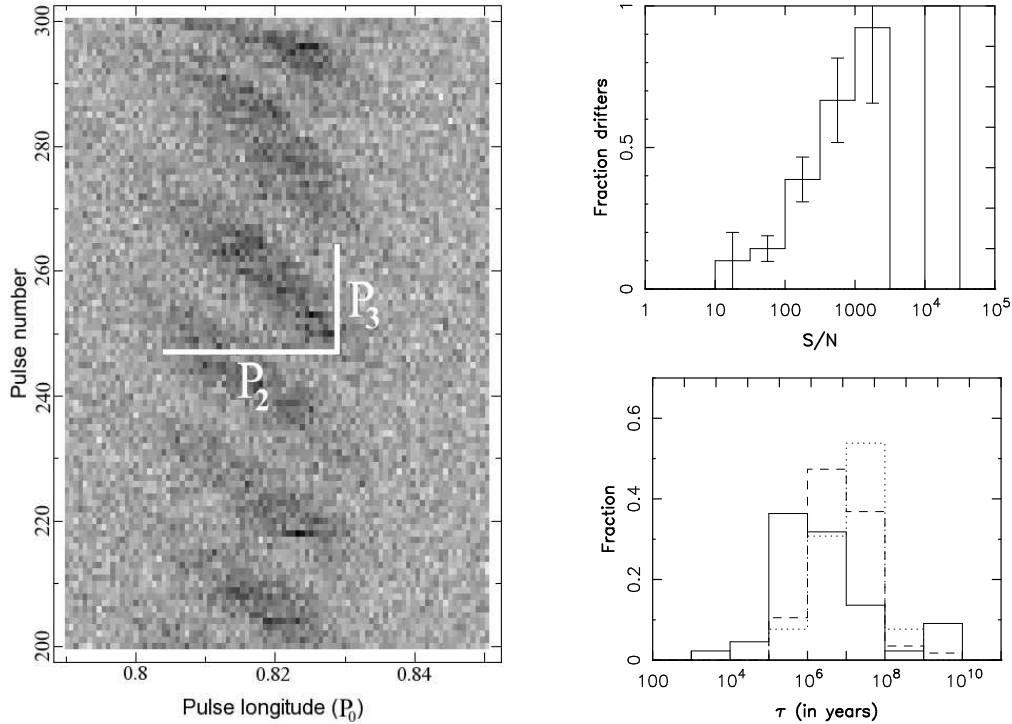}}}
\end{minipage}
\begin{minipage}{.48\textwidth}
\begin{center}
\rotatebox{270}{\resizebox{!}{0.8\hsize}{\includegraphics[angle=0]{s2n_hist2.ps}}}\\\vspace{3mm}
\rotatebox{270}{\resizebox{!}{0.77\hsize}{\includegraphics[angle=0]{age_hist2.ps}}}
\end{center}
\end{minipage}
\end{center}
\caption{\label{Classes_fig}{\bf Left:} A pulse-stack of the new
drifter PSR B1819$-$22. \label{s2n_hist}{\bf Top right:} The fraction
of pulsars we observe to show the drifting phenomenon versus the
measured $S/N$ ratio of the observation.
{\bf Bottom right:}\label{age_hist} The age distribution of the
non-drifting pulsars (solid line), all the drifters (dashed line) and
the coherent drifters (dotted line).
}
\end{figure}

If one can detect single pulses one can see that in some pulsars they
consist of subpulses and in some cases these subpulses drift in
successive pulses in an organized fashion through the pulse window. If
one plots a so-called ``pulse-stack'', a plot in which successive
pulses are displayed on top of one another, the drifting phenomenon
causes the subpulses to form ``drift bands''. In the left panel of
Fig. \ref{Classes_fig} one can see a sequence of 100 pulses of one of
the new drifters we have found which clearly shows the drifting
phenomenon.  The pulse number is plotted vertically and the time
within the pulses (i.e. the pulse longitude) horizontally. The drift
bands are characterized by two numbers: the horizontal separation
between them in pulse longitude ($P_2$) and the vertical separation in
pulse periods ($P_3$).  This complex, but highly regular intensity
modulation in time is known in great detail for only a small number of
well studied pulsars. Because the properties of the subpulses are most
likely determined by the emission mechanism, we learn about the
physics of the emission mechanism by studying them. That drifting is
linked to the emission mechanism is suggested by the fact that
drifting is affected by ``nulls'', where nulling is the phenomenon
whereby the emission mechanism switches off for a number of successive
pulses. The main goals of this unbiased search for pulsar subpulse
modulation is to determine what percentage of the pulsars show the
drifting phenomenon and to find out if these drifters share some
physical properties. As a bonus of this observational program new,
individually interesting drifting, subpulse systems are found
(Weltevrede et~al.~\cite{wes06}).


\section{Observations and data analysis}
An important aspect when calculating the statistics of drifting is
that one has to be as unbiased as possible, so we have selected our
sample of pulsars based only on the predicted signal-to-noise ($S/N$)
ratio in a reasonable observing time. While this sample is obviously
still luminosity biased, it is not biased towards well-studied
pulsars, pulse profile morphology or any particular pulsar
characteristics as were previous studies (e.g.  Ashworth~\cite{ash82},
Backus~\cite{bac81} and Rankin~\cite{ran86}). Moreover, all the
conclusions in this paper are based on observations at a single
frequency.  All the analyzed observations were collected with the WSRT
in the Netherlands at an observation wavelength of 21 cm.

One basic method to find out if there is subpulse modulation is to
calculate the modulation index, which is a measure of the factor by
which the intensity varies from pulse to pulse and could therefore be
an indication for the presence of subpulses. To determine if the
subpulses are drifting, the Two-Dimensional Fluctuation Spectrum
(2DFS; Edwards \& Stappers~\cite{es02}) is calculated. By analyzing
the 2DFS it can be determined if this modulation is disordered or
(quasi-)periodic and if there exists a systematic drift. The
calculation of the 2DFS is an averaging process and this makes it a
powerful tool to detect drifting subulses, even when the $S/N$ is too
low to detect single pulses.  The drifting is classified as {\em
coherent} when the drift has a well defined $P_3$ value. For more
details about the observations and data analysis we refer to
Weltevrede et~al. (\cite{wes06}).

\section{Statistics}

\subsection{The Numbers}

Our sample of pulsar is not biased on pulsar type or any particular
pulsar characteristics. This allows us, first of all, to address the
very basic question: what fraction of the pulsars show the drifting
phenomenon?  Of the \NrPulsars analyzed pulsars \NrDrifters pulsars
show the drifting phenomenon, indicating that at least one in three
pulsars drift. This is a lower limit for a number of reasons. First of
all, not all the observations have the expected $S/N$. This could be
because of radio interference, interstellar scintillation,
digitization effects, or because the flux or pulse width for some
pulsars was wrong in the database used.

In the top right panel of Fig. \ref{s2n_hist} the fraction of pulsars
that show the drifting phenomenon is plotted versus the $S/N$ ratio of
the observation. One can see that the probability of detecting
drifting is higher for observations with a higher $S/N$. To make the
statistics more independent of the $S/N$ ratio of the observations,
the statistics are done with the \NrPulsarsSNR pulsars with a $S/N
\geq 100$. Of these pulsars \DriftPercentageSNR\% is detected to be
drifters and from the top right panel of Fig. \ref{s2n_hist} it is
clear that the real drift percentage could even be higher.  There are
many reasons why drifting is not expected to be detected for all
pulsars. For instance for some pulsars the line of sight cuts the
magnetic pole centrally and therefore longitude stationary subpulse
modulation is expected. Also, refractive distortion in the pulsar
magnetosphere or nulling will disrupt the drift bands, making it
difficult or even impossible to detect drifting. Some pulsars are
known to show organized drifting subpulses in bursts. In that case (or
when $P_3$ is very large) some of our observations could be too short
to detect the drifting.

With a lower limit of one in two it is clear that drifting is at the
very least a common phenomenon for radio pulsars.  This is consistent
with the conclusion that the drifting phenomenon is only weakly
correlated with (or even independent of) magnetic field strength
(Weltevrede et~al.~\cite{wes06}), because the drifting phenomenon is
too common to require very special physical conditions. It could well
be that the drifting phenomenon is an intrinsic property of the
emission mechanism although for some pulsars it is difficult or even
impossible to detect.

\subsection{The Age Dependence Of The Drifting Phenomenon}

Two directly measurable and therefore important physical parameters of
the pulsar are the pulse period and its time derivative (spin-down
parameter). From the pulsar age histograms (bottom right panel of
Fig. \ref{age_hist}) it can be seen that the population of pulsars
that show the drifting phenomenon is on average older than the
population of pulsars that do not show drifting. Moreover it seems
that drifting is more coherent for older pulsars.  It turns out that
the drifters and nondrifters have significantly different age
distributions and that the pulsars which drift coherently are likely
to have a separate age distribution (Weltevrede et~al.~\cite{wes06}).
It is intriguing to think that drifting becomes more and more coherent
for pulsars with a higher age. A possible mechanism to distort the
drift bands is nulling. However it has been found that the nulling
fraction is on average higher for older pulsars, showing that nulling
cannot explain this correlation.

Another possible scenario is that the alignment of the magnetic dipole
axis with the rotation axis has something to do with the observed
trend. Observations seem to show that the angle $\alpha$ between the
magnetic axis and the rotation axis is on average smaller for older
pulsars and this angle is likely to be an important physical parameter
in the mechanism that drives the drifting phenomenon.  In this
scenario as the pulsar gets older, the rotation axis and the magnetic
axis grows more aligned, which makes the drifting mechanism more
effective or regular.  Also the pulse profile morphology seems to
evolve when the pulsar ages what could make drifting subpulses more
likely to be detected in older pulsars. In the non-radial pulsations
model this trend can also be explained, because the appearance of
narrow drifting subpulses is favored in pulsars with an aligned
magnetic axis (Clemens \& Rosen~\cite{cr04}).

\subsection{The Drifting Phenomenon And The Modulation Index}
\label{ModulationSection}

\begin{figure}[tb]
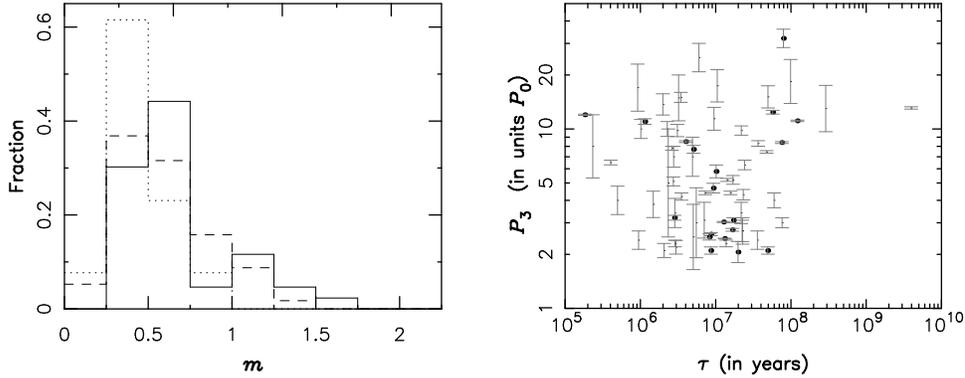

\begin{center}
\rotatebox{270}{\resizebox{!}{0.4\hsize}{\includegraphics[angle=0]{minmod_hist2.ps}}}\hspace{0.05\hsize}
\rotatebox{270}{\resizebox{!}{0.42\hsize}{\includegraphics[angle=0]{p3_age2.ps}}}
\end{center}
\caption{{\bf Left:} \label{mod_hist} The modulation index
distribution of the pulsars that do not show the drifting phenomenon
(solid line), that do show the drifting phenomenon (dashed line) and
of the pulsars that drift coherently (dotted line).  {\bf Right:}
\label{p3age} The measured value of the vertical drift band separation
$P_3$ versus the pulsar age. The coherent drifters are the filled
circles.  }
\end{figure}

The drifting phenomenon is a form of subpulse modulation, so the
modulation index is an obvious parameter to try to correlate with the
drifting phenomenon.  Modulation index distributions are shown in the
left panel of Fig. \ref{mod_hist}. Readily apparent is the trend that
pulsars that show the drifting phenomenon more coherently have on
average a lower modulation index (not shown to be statistically
significant).

To explain the trend, pulsars that drift coherently must either have
on average more subpulses per pulse or the subpulse intensity
distribution must be more narrow.  The latter could be understood
because coherent drifting could indicate that the electrodynamical
conditions in the sparking gap are stable. Also the presence of
subpulse phase steps results in a lower modulation index and could be
explained as the result of interference between two superposed
drifting subpulse signals that are out of phase (e.g. Edwards \&
Stappers~\cite{es03c}). It is not unlikely that this interference can
only occur if the drifting is coherent, which could explained the
trend. It is also found that many pulsars must have a non-varying
component in their emission, consistent with the presence of
superposed out of phase subpulse signals.  Another explanation for
this trend would be that for some pulsars the organized drifting
subpulses are more refractively distorted than for others, causing the
subpulses to appear more disordered in the pulse window. Moreover it
could be expected that the intensities of the individual subpulses
varies more because of lensing (e.g. Petrova~\cite{pet00}) and
possible focusing of the radio emission (Weltevrede
et~al.~\cite{wsv+03}), causing the modulation index to be higher in
those pulsars.

The modulation index of core type emission is observed to be in
general lower than that of conal type of emission. This is also a
consequence of the Gil \& Sendyk (\cite{gs00}) model. In the sparking
gap model, the drifting phenomenon is associated with conal emission
and therefore expected to be seen in pulsars with an on average higher
modulation index. If well organized coherent drifting is an
exclusively conal phenomenon, it is expected that coherent drifters
have an on average a higher modulation index, exactly opposite to the
observed trend. No drifting is expected for pulsars classified as
``core single stars''. Although this may be true for many cases there
are some exceptions, stressing the importance of being unbiased on
pulsar type when studying the drifting phenomenon.

In the framework of the sparking gap model the subpulses are generated
(indirectly) by discharges in the polar gap (i.e. sparks). The number
of sparks that fits on the polar cap is quantified by the complexity
parameter (Gil \& Sendyk~\cite{gs00}), which is expected to be
anti-correlated with the modulation index (Jenet \&
Gil~\cite{jg03}). The complexity parameter is a function of the pulse
period and its derivative and its precise form depends on the model
one assumes for the pulsar emission. By correlating the modulation
index of a sample of pulsars with various complexity parameters as
predicted by different emission models one could try to distinguish
which model best fits the data. We have correlated the modulation
indices in our sample of pulsars with the complexity parameter of four
different emission models as derived by Jenet \& Gil
(\cite{jg03}). Unfortunately none of the models can be ruled out based
on these observations.

\subsection{Properties Of The Drift Behavior}

A significant correlation between $P_3$ and the pulsar age has been
reported in the past (e.g. Rankin~\cite{ran86}). As one can see in the
right panel of Fig. \ref{p3age} there is no clear correlation found in
our data, which is confirmed by $\chi^2$-fitting. There is no
correlation found between $P_3$ and the magnetic field strength or the
pulse period as well as between the drift direction and the pulsar
spin-down as reported in the past (Ritchings \& Lyne~\cite{rl75}).
The evidence for a pulsar subpopulation located close to the
$P_3=2P_0$ Nyquist limit also seems to be weak.

In a sparking gap model one would expect that the spark-associated
plasma columns drift because of an $\mathbf{E}\times \mathbf{B}$
drift, which depends on both the pulse period and its derivative.  The
absence of any correlation between $P_3$ and a physical pulsar
parameter is difficult to explain in this model, unless many pulsars
in our sample are aliased. Because the emission entities are only
sampled once per rotation period of the star, it is very difficult to
determine if the subpulses in one drift band correspond to the same
emission entity for successive pulses. For instance for PSR B1819$-$22
(see left panel of Fig. \ref{Classes_fig}) we do not know if the
emission entities drift slowly toward the leading part of the pulse
profile (not aliased) or faster toward the trailing part of the pulse
profile (aliased).  If a pulsar is aliased a higher $\mathbf{E}\times
\mathbf{B}$ drift can result in a lower $P_3$ value and visa versa,
making $P_3$ not a direct measure of the $\mathbf{E}\times \mathbf{B}$
drift.  Also if $P_2$ is highly variable from pulsars to pulsar, any
correlation with $P_3$ is expected to be weaker.

\section{Conclusions}
The number of pulsars that are known to show the drifting phenomenon
is significantly expanded by 42 and the fraction of
pulsars that show the drifting phenomenon is likely to be larger than
55\%. This implies that the physical conditions required for the
drifting mechanism to work cannot be very different than the required
physical conditions for the emission mechanism of radio pulsars. It
could well be that the drifting phenomenon is an intrinsic property of
the emission mechanism, although drifting could in some cases be very
difficult to detect.

Our results seem to suggest that drifting is not exclusively related
to conal emission. Our sample of pulsars is not biased on pulsar type
or any particular pulsar characteristics, which allows us to do
meaningful statistics on the drifting phenomenon.  Although
significant correlations between $P_3$ and the pulsar age, the
magnetic field strength and the pulse period have been reported, we
find no such correlations in our enlarged sample.
The absence of a correlation between $P_3$ and any physical pulsar
parameter is difficult to explain, unless many pulsars in our sample
are aliased or if $P_2$ is highly variable from pulsar to pulsar.

The population of pulsars that show the drifting phenomenon are on
average older than the population of pulsars that do not show drifting
and it seems that drifting is more coherent for older pulsars. The
evolutionary trend found seems to suggest that the mechanism that
generates the drifting subpulses gets more and more stable as the
pulsar ages.

If subpulse phase steps are exclusively (or at least more likely) to
occur in pulsars with coherently drifting subpulses, their modulation
index is expected to be on average lower. This is indeed the trend the
we observe. Another possible scenario to explain this trend is that
coherent drifting indicates that the electrodynamical conditions in
the sparking gap are stable or that refraction in the magnetosphere is
stronger for pulsars that do not show the drifting phenomenon
coherently.


\label{lastpage}

\end{document}